\documentclass[fleqn,usenatbib,useAMS]{mnras}

\usepackage{mathptmx}

\usepackage[T1]{fontenc}
\usepackage{ae,aecompl}

\usepackage{graphicx}	
\usepackage{amsmath}	
\usepackage{amssymb}	
\usepackage{bm}		
\usepackage{pdflscape}	
\usepackage{color}
\usepackage{xcolor}
\usepackage{exscale}
\usepackage{natbib}
\usepackage{rotating}
\usepackage{rotate}
\usepackage{afterpage}
\usepackage{booktabs,threeparttable}
\usepackage{relsize}
\usepackage{lscape}
\usepackage{tabularx}
\usepackage{longtable}
\usepackage{ltxtable}
\usepackage{txfonts}
\usepackage{bbm}
\usepackage{widetext}
\usepackage{units,multirow}
\usepackage{xurl}
\usepackage{hyperref}

\usepackage{soul}

\newcommand{\ie}{{i.e.}}

\newcommand{\eg}{{e.g.}}

\newcommand{\bv}[1]{\mathbfit{#1}}
\newcommand{\bmat}[1]{\mathbfss{#1}}
\newcommand{\x}{$\times$}

\newcommand{\mr}{\mathrm}

\newcommand{\tbf}{\textbf}
\newcommand{\matC}{\bmat{C}}

\newcommand{\like}{L}

\newcommand{\om}{\Omega_\mr m}
\newcommand{\omb}{\Omega_\mr b}
\newcommand{\sig}{\sigma_8}
\newcommand{\ns}{n_s}
\newcommand{\w}{w_0}
\newcommand{\wa}{w_a}
\newcommand{\D}{\bv{D}}
\newcommand{\M}{\bv{M}}

\newcommand{\pco}{\bv{p}_{\rm co}}
\newcommand{\pnu}{\bv{p}_{\rm nu}}

\newcommand{\be}{\begin{equation}}
\newcommand{\ee}{\end{equation}}
\newcommand{\bea}{\begin{eqnarray}}
\newcommand{\eea}{\end{eqnarray}}

\title[Synergies between Roman ST and CMB experiments]{Cosmology from weak lensing, galaxy clustering, CMB lensing and tSZ:\\
II. Optimizing Roman survey design for CMB cross-correlation science}

\author[Eifler et al.]{Tim Eifler\thanks{timeifler@arizona.edu}$^{1,2}$, Xiao Fang$^{1,3}$, Elisabeth Krause$^{1,2}$, Christopher M. Hirata$^{4,5,6}$, Jiachuan Xu$^{1}$,
\newauthor
 Karim Benabed$^{7}$, Simone Ferraro$^{8,3}$, Vivian Miranda$^{9}$, Pranjal R. S.$^{1}$, Emma Ayçoberry$^{7}$, Yohan Dubois$^{7}$\\
$^{1}$Department of Astronomy and Steward Observatory, University of Arizona, 933 North Cherry Avenue, Tucson, AZ 85721, USA\\
$^{2}$Department of Physics, University of Arizona, 1118 E. Fourth Street, Tucson, AZ 85721, USA\\
$^{3}$Berkeley Center for Cosmological Physics, UC Berkeley, CA 94720, USA\\
$^{4}$Center for Cosmology and AstroParticle Physics (CCAPP), The Ohio State University, 191 West Woodruff Ave, Columbus, OH 43210, USA\\
$^{5}$Department of Physics, The Ohio State University, 191 West Woodruff Ave, Columbus, OH 43210, USA\\
$^{6}$Department of Astronomy, The Ohio State University, 140 West 18th Avenue, Columbus, OH 43210, USA\\
$^{7}$Sorbonne Université, CNRS, UMR 7095, Institut d’Astrophysique de Paris, 98 bis bd Arago, 75014 Paris, France\\
$^{8}$Lawrence Berkeley National Laboratory, One Cyclotron Road, Berkeley, CA 94720, USA\\
$^{9}$C. N. Yang Institute for Theoretical Physics, Stony Brook University, Stony Brook, NY 11794
}
\date{Accepted XXX. Received YYY; in original form ZZZ}

\pubyear{2023}

\begin{document}
\label{firstpage}
\pagerange{\pageref{firstpage}--\pageref{lastpage}}
\maketitle


 \begin{abstract} 
We explore synergies between the Nancy Grace Roman Space Telescope High Latitude Wide Area Survey (HLWAS) and CMB experiments, specifically Simons Observatory (SO) and CMB-Stage4 (S4). Our simulated analyses include weak lensing, photometric galaxy clustering, CMB lensing, thermal SZ, and cross-correlations between these probes. While we assume the nominal 16,500 deg$^2$ area for SO and S4, we consider multiple survey designs for Roman that overlap with Rubin Observatory's Legacy Survey of Space and Time (LSST): the 2000 deg$^2$ reference survey using four photometric bands, and two shallower single-band surveys that cover 10,000 and 18,000 deg$^2$, respectively. We find a $\sim2\times$ increase in the dark energy figure of merit when including CMB-S4 data for all Roman survey designs. We further find a strong increase in constraining power for the Roman wide survey scenario cases, despite the reduction in galaxy number density, and the increased systematic uncertainties assumed due to the single band coverage. Even when tripling the already worse systematic uncertainties in the Roman wide scenarios, which reduces the 10,000 deg$^2$ FoM from 269 to 178, we find that the larger survey area is still significantly preferred over the reference survey (FoM 64). We conclude that for the specific analysis choices and metrics of this paper, a Roman wide survey is unlikely to be systematics-limited (in the sense that one saturates the improvement that can be obtained by increasing survey area). We outline several specific implementations of a two-tier Roman survey (1000 deg$^2$ with 4 bands, and a second wide tier in one band) that can further mitigate the risk of systematics for Roman wide concepts. 

\end{abstract}


\renewcommand{\thefootnote}{\arabic{footnote}}
\setcounter{footnote}{0}
\section{Introduction}
\label{sec:intro}
Observing the large-scale structure in our Universe provides a wealth of cosmological information that allows us to constrain fundamental physics questions such as the nature of dark energy, the mass and number of species of neutrinos, possible modifications to General Relativity as a function of scale or environment, or the nature of dark matter interactions. 

Early results from the Dark Energy Survey \citep[DES\footnote{www.darkenergysurvey.org/},][]{Y3methods,Y33x2,PandeyY3}, the Kilo-Degree Survey \citep[KiDS\footnote{http://www.astro-wise.org/projects/KIDS/},][]{KIDS1000}, and the Hyper Suprime Cam Subaru Strategic Program \citep[HSC\footnote{http://www.naoj.org/Projects/HSC/HSCProject.html},][]{MiyatakeHSCY3,SugiyamaHSCY3} have shown promising results of multi-probe analyses from photometric data. 

The advent of data from Stage 4 surveys will significantly increase the scientific discovery potential \citep[see][for a review]{2013PhR...530...87W}. In terms of photometric surveys, the Rubin Observatory Legacy Survey of Space and Time \citep[LSST\footnote{https://www.lsst.org/},][]{2019ApJ...873..111I}, Euclid\footnote{https://sci.esa.int/web/euclid} \citep{2011arXiv1110.3193L}, the Spectro-Photometer for the History of the Universe, Epoch of Reionization, and Ices Explorer \citep[SPHEREx\footnote{http://spherex.caltech.edu/},][]{2014arXiv1412.4872D}, and \textit{Roman} \citep[Roman Space Telescope\footnote{https://roman.gsfc.nasa.gov/},][]{2015arXiv150303757S} will jolt the community into a new regime of data volume and accuracy. 

These photometric datasets are highly synergistic with spectroscopic datasets e.g., from the Dark Energy Spectroscopic Instrument \citep[DESI, ][]{2016arXiv161100036D}, the Prime Focus Spectrograph \citep[PFS, ][]{2014PASJ...66R...1T}, and the spectroscopic components of Euclid and Roman. Together with the next generation of CMB experiments, e.g., the Simons Observatory \citep[SO, ][]{2019JCAP...02..056A} and CMB-S4 \citep{2016arXiv161002743A} the cosmology community will have access to a large set of accurate and complementary datasets to explore fundamental physics.

Synergies of galaxy weak lensing and clustering with CMB lensing has been studied extensively in the context of upcoming galaxy surveys and CMB experiments \citep[\eg,][]{2017PhRvD..95l3512S,2020JCAP...12..001S,2018PhRvD..97l3540S, 2022MNRAS.509.5721F,2022MNRAS.512.5311W}, and the synergies have been carried out in experiments, including SDSS + Planck \citep[\eg,][]{2017MNRAS.464.2120S}, DES + Planck \citep[\eg,][]{2016MNRAS.456.3213G,2019PhRvD.100d3501O,2019PhRvD.100d3517O,2019PhRvD.100b3541A}, DES + Planck + SPT \citep[\eg][]{2023PhRvD.107b3529O,2023PhRvD.107b3530C,2023PhRvD.107b3531A}, unWISE + Planck \citep[\eg,][]{2021JCAP...12..028K}, HSC + Planck \citep[\eg,][]{2022PhRvL.129f1301M}, KiDS + ACT + Planck \citep[\eg,][]{2021A&A...649A.146R}, DESI / DESI-like + Planck \citep[\eg,][]{2021MNRAS.501.1481H,2021MNRAS.501.6181K,2022JCAP...02..007W}. 

More recently, the thermal Sunyaev Zel'dovich effect has been identified as a promising additional source of information, especially in the context of self-calibrating baryonic effects in galaxy weak lensing. Several important contributions to the literature explore these synergies in both theory \citep[\eg,][]{2020PhRvD.101d3525P,2022JCAP...04..046N} and in observations, such as Canada France Hawaii Lensing Survey + Planck \citep{2015JCAP...09..046M}, 2MASS + Planck \citep{2018MNRAS.480.3928M}, 2MASS + WISE\x SuperCOSMOS + Planck \citep{2020MNRAS.491.5464K}, SDSS + Planck \citep[\eg,][]{2017MNRAS.467.2315V,2018PhRvD..97h3501H,2020MNRAS.491.2318T,2020ApJ...902...56C}, SDSS + ACT \citep[\eg,][]{2021PhRvD.103f3513S}, KiDS + Planck + ACT \citep[\eg,][]{2021A&A...651A..76Y,2022A&A...660A..27T}, and DES + Planck + ACT \citep[\eg,][]{2022PhRvD.105l3525G,2022PhRvD.105l3526P}.
 
This paper is the second in a series of two where the first paper \citep{2024MNRAS.527.9581F} develops the methodology for a 10\x2 analysis that combines weak lensing, galaxy clustering, CMB lensing, and tSZ auto-correlations and their cross-correlations. This second paper explores the 10\x2 concept in the context of the Roman Space Telescope and CMB experiments (S4 and SO) with a particular focus on optimising the survey strategy for the imaging component of Roman's High Latitude Wide Area Survey.

In its \textit{reference design} the HLWAS assumes a duration of $\sim$1 year covering 2,000 deg$^2$ with the four-band imaging. Several alternative designs have been suggested in \cite{esk21} e.g., to maximise synergies with LSST. The trade space of maximising the statistical power of Roman while maintaining exquisite systematics control is one of the most important near-term challenges in Roman survey design.

In order to answer these questions precision forecasts that accurately reflects the likelihood analyses that will be conducted on the actual data are necessary. Given the proximity of the launch date and given the experience that the community has with precursor data, we are now in a position to simulate such Roman likelihood analyses over a wide range of science cases, survey strategies, systematics scenarios, including synergies with external datasets. Of particular interest in this context is the idea of a two-tier Roman HLWAS, where the reference design survey mode is only used over 1000 deg$^2$ and the remaining time is used on a single-band wide survey. 

This paper is a contribution to these efforts and structured as follows: We describe the theoretical basics that underpin our modelling calculations in Sect.~\ref{sec:basics}, before describing our survey specific modelling of systematics and statistical uncertainties and related inference techniques for the simulated analyses (Sect.~\ref{sec:like}). We continue exploring synergies between Roman HLWAS and S4/SO using the reference design (Sect.~\ref{sec:RomanCMB}) and a Roman wide survey design that covers the full area of SO/S4 (Sect.~\ref{sec:wide}). Here, we also explore different levels of observational and astrophysical systematics and we quantify the constraining power when using a different lens sample in the 10\x2 analysis, namely the source sample itself. We continue in Sect. \ref{sec:wide-designs} with specific implementations of wide-survey concepts, in particular when combining the Roman wide survey with a reduced footprint that is covered in the reference design surveymode. We conclude in Sect.~\ref{sec:conc}.

\section{Theoretical Concepts}
\label{sec:basics}
We summarise the underlying theoretical equations that enter our modelling in this section. For a more detailed derivation of some of the equations the reader is referred to \cite{2024MNRAS.527.9581F}.

\subsection{Projected Power Spectra}\label{ssec:probe-modeling}
We use capital Roman subscripts to denote observables, $A,B\in \left\{\kappa_\mr{g},\delta_\mr{g}, \kappa_\mr{CMB}, y \right\}$, where $\kappa_\mr{g}$ refers to the lensing of source galaxies, $\delta_{\mathrm{g}}$ is the density contrast of the lens galaxy sample, $\kappa_\mr{CMB}$ is the CMB lensing map, and $y$ is the Compton y parameter given by the integrated electron pressure along the line of sight, which traces the tSZ effect. 

All angular power spectrum between redshift bin $i$ of observable $A$ and redshift bin $j$ of observables $B$ at projected harmonic mode $\ell$, $C_{AB}^{ij}(\ell)$ are computed using the Limber \citep{1992ApJ...388..272K} and flat sky approximations \citep[see][for impact of these approximations on actual data analyses]{2020JCAP...05..010F}:
\begin{equation}
\label{eq:projected}
C_{AB}^{ij}(\ell) = \int d\chi \frac{q_A^i(\chi)q_B^j(\chi)}{\chi^2}P_{AB}(k=\ell/\chi,\;z(\chi)) \,.
\end{equation}
We denote $\chi$ as the comoving distance, $q_A^i(\chi)$ are weight functions of the different observables given in Eqs.~(\ref{eq:q1}-\ref{eq:q4}). The 3D cross-power spectra $P_{AB}(k,z)=P_{BA}(k,z)$ can be related to the nonlinear matter power spectrum $P_{\delta\delta}(k,z)$, matter-electron pressure power spectrum $P_{\delta P_e}(k,z)$, and electron pressure power spectrum $P_{P_eP_e}(k,z)$ by the following relations: 
\begin{align}
    &P_{\delta_{\rm g}B}(k,z) = b_{\rm g}(z)P_{\delta B}(k,z)~, \label{eq:3d1} \\
    &P_{\kappa_{\rm g}B}(k,z) = P_{\kappa_{\rm CMB}B}(k,z) = P_{\delta B}(k,z)~, \label{eq:3d2}\\
    &P_{yB}(k,z)= P_{P_e B}(k,z)~, \label{eq:3d3}
\end{align}
where $\delta$ is the nonlinear matter density contrast and $P_{\rm e}$ is the electron pressure. Due to Eq.~(\ref{eq:3d3}), we will use $P_{yB}$ to represent $P_{P_eB}$, which simplifies the subscripts. Thus, $P_{\delta P_e}\equiv P_{\delta y}$ and $P_{P_e P_e}\equiv P_{yy}$. We calculate $P_{AB}$ ($A,B\in\lbrace\delta,y\rbrace$) using a halo model implementation based on the HMx parameterisation from \cite{2020A&A...641A.130M}, presented in Sect.~\ref{ssec:halo-model}.

Throughout this paper we assume that the galaxy density contrast is proportional to the nonlinear matter density contrast, fully described by an effective galaxy bias parameter $b_{\rm g}(z)$. This assumption breaks down on small scales \citep[e.g.,][]{2017MNRAS.470.2100K,2017JCAP...08..009M,2020PhRvD.102l3522P,2022PhRvD.106d3520P, 2021JCAP...12..028K,2021MNRAS.501.1481H,2021MNRAS.501.6181K}; consequently we choose conservative scale cuts on probes that include $\delta_{\rm g}$ related. 

The weight function for the projected power spectra $q_{A}^i(\chi),q_{B}^j(\chi)$ of the observables $A,B$ in redshift bin $i,j$ are given by
\begin{align}
    &q_{\delta_{\rm g}}^i(\chi) = \frac{n_{\rm lens}^i(z(\chi))}{\bar{n}^i_{\rm lens}}\frac{dz}{d\chi}~, \label{eq:q1}\\
    &q_{\kappa_{\rm g}}^i(\chi) = \frac{3H_0^2\Omega_m}{2c^2}\frac{\chi}{a(\chi)}\int_{\chi_{\rm min}^i}^{\chi_{\rm max}^i}\,d\chi'\frac{n_{\rm source}^i(z(\chi'))}{\bar{n}_{\rm source}^i}\frac{dz}{d\chi'}\frac{\chi'-\chi}{\chi'}~,\label{eq:q2} \\
    &q_{\kappa_{\rm CMB}}(\chi) = \frac{3H_0^2\Omega_m}{2c^2}\frac{\chi}{a(\chi)}\frac{\chi^*-\chi}{\chi^*}~, \label{eq:q3}\\
    &q_y(\chi) = \frac{\sigma_{\rm T}}{m_e c^2}\frac{1}{[a(\chi)]^2}~, \label{eq:q4}
\end{align}
where $\chi_{\rm min/max}^i$ is the minimum / maximum comoving distance of the redshift bin $i$, $a(\chi)$ is the scale factor, $\Omega_m$ the matter density fraction at present, $H_0$ the Hubble constant, $c$ the speed of light, $\chi^*$ the comoving distance to the surface of last scattering, $\sigma_{\rm T}$ the Thomson scattering cross section, and $m_e$ the electron mass.

\subsection{3D Power Spectra - Halo Model}
\label{ssec:halo-model}
We write the 3D power spectrum of two fields $A$ and $B$ as the sum of the 2-halo term (the 1st term below) and the 1-halo term (the 2nd),
\begin{equation}
    P_{AB}(k,z) = P_{\rm lin}(k,z)I_{A}^1(k,z)I_{B}^1(k,z) + I_{AB}^0(k,k,z)~,
\end{equation}
where $P_{\rm lin}(k,z)$ is the linear matter power spectrum computed by the \cite{1998ApJ...496..605E} fitting formula. Following \cite{2024MNRAS.527.9581F} notation\footnote{
For given halo mass function weighted quantity $X(M)$
\begin{equation}
    \langle X(M)\rangle_{\scriptscriptstyle M} \equiv \int_{M_{\rm min}}^{M_{\rm max}} dM\frac{dn(M)}{dM}X(M)~,
\end{equation}
where $M$ here denotes halo mass and $dn(M)/dM$ is the halo mass function.} we express
\begin{align}
    I_{A}^\alpha(k,z) &= \langle b_{h,\alpha}(M,z) \tilde{u}_{A}(k,M,z)\rangle_{\scriptscriptstyle M}~, \label{eq:I1}\\
    I_{AB}^\alpha(k,k',z) &= \langle b_{h,\alpha}(M,z) \tilde{u}_{A}(k,M,z)\tilde{u}_{B}(k',M,z)\rangle_{\scriptscriptstyle M}~, \label{eq:I2}
\end{align}
where $b_{h,\alpha}$ is the $\alpha$-th order halo biasing, with $b_{h,0}=1$. We employ the \cite{2010ApJ...724..878T} fitting function to compute the halo bias $b_{h,1}$, and the \cite{2008ApJ...688..709T} fitting function for the halo mass function $dn/dM$. The functions $\tilde{u}_X(k,M,z)$ ($X\in\lbrace \delta,y\rbrace$) are related to the Fourier transforms of the radial profiles of matter density $\rho_m(M,r,z)$ and electron pressure $P_e(M,r,z)$ within a halo of mass $M$. We assume the NFW halo profile \citep{1997ApJ...490..493N} to model $\tilde{u}_\delta(k,M,z)$.

For the baryon distribution characterised through the $y$-field we closely follow the HMx treatment \citep{2020A&A...641A.130M}, and assume that the gas has a bound and ejected component. The former is found in the 1-halo and 2-halo term whereas the latter contributes to the 2-halo term only. 
\begin{align}
    I_y^1(k,z)&=\langle b_{h,1}(M,z)(\tilde{u}_{y,\rm bnd}(k,M,z) +\tilde{u}_{y,\rm ejc}(M,z))\rangle_{\scriptscriptstyle M}~, \label{eq:I3} \\
    I_{y\delta}^0(k,k,z)&=\langle \tilde{u}_{y,\rm bnd}(k,M,z)\tilde{u}_{\delta}(k,M,z)\rangle_{\scriptscriptstyle M}~, \label{eq:I4} \\
    I_{yy}^0(k,k,z)&=\langle \tilde{u}_{y,\rm bnd}(k,M,z)\tilde{u}_{y,{\rm bnd}}(k,M,z)\rangle_{\scriptscriptstyle M}~ \label{eq:I5} .
\end{align}
The two components are computed assuming that bound gas follows the Komatsu-Seljak profile \citep{2002MNRAS.336.1256K}, while the ejected gas follows the linear perturbations of the matter field. 

The exact derivation can be found in \cite{2024MNRAS.527.9581F}, the final expressions read:
\begin{equation}
    \tilde{u}_{y,\rm ejc}(M) = \frac{Mf_{\rm ejc}(M)}{m_{\rm p}\mu_e}k_B T_{\rm w}~
\end{equation}
\begin{align}
    \tilde{u}_{y,\rm bnd}(k,M)=\frac{2\alpha}{3}\frac{GM^2f_{\rm bnd}(M)\mu_{\rm p}}{a r_v\mu_e}\frac{F(c(M),kr_s)}{F_0(c(M))} ~,
\end{align}
where $\mu_{\rm p} = 4/(3 + 5 f_{\rm H})$, $\mu_e=2/(1+f_{\rm H})$, $f_{\rm H}$ is the hydrogen mass fraction, $k_B$ is the Boltzmann constant, and $T_{\rm w}$ is the ejected diffuse warm gas temperature. The halo scale radius parameter $r_s=r_{\rm v}/c(M,z)$, where $r_{\rm v}$ satisfies $M=4\pi r_{\rm v}^3\bar{\rho}\Delta_{\rm v}/3$, $\bar{\rho}$ is the comoving mean matter density, $\Delta_{\rm v}\bar{\rho}$ is the virial-collapse density, and $c(M,z)$ is the mass-concentration relation. We assume $\Delta_{\rm v}=200$. Parameter $\alpha$ encapsulates deviations from a simple virial relation and we take $\alpha=1$ by default. The fractions of bound and ejected gas in halos are modelled as
\begin{align}
    f_{\rm bnd}(M)&=\frac{\Omega_{\rm b}}{\Omega_{\rm m}}\frac{1}{1+(M_0/M)^\beta}~,\label{eq:fbnd}\\
    f_{\rm ejc}(M)&=\frac{\Omega_{\rm b}}{\Omega_{\rm m}}-f_{\rm bnd}(M)-A_* \exp\left[-\frac{\log_{10}^2(M/M_*)}{2\sigma_*^2}\right]~, \label{eq:fejc}
\end{align}
where parameters $M_0,\beta$ describe the effect that the fraction of ejected gas increases as the halo mass decreases for halo mass below a critical value $M_0$, and the stellar fraction is parameterised by $A_*, M_*, \sigma_*$, which are fixed at fiducial values as explained in \cite{2024MNRAS.527.9581F}. Functions $F(c,y)$ and $F_0(c)$ are defined as 
\begin{align}
    &F(c,y)=\int_0^{c} dx\,x^2j_0(xy)\left[\frac{\ln(1+x)}{x}\right]^{\Gamma/(\Gamma-1)}~, \\
    &F_0(c)=\int_0^{c} dx\,x^2\left[\frac{\ln(1+x)}{x}\right]^{1/(\Gamma-1)}~,
\end{align}
where parameter $\Gamma$ is the polytropic index for the gas. The integrals must be solved numerically. We use a public FFTLog code\footnote{\url{https://github.com/xfangcosmo/FFTLog-and-beyond}} \citep{2020JCAP...05..010F} to solve the Bessel integral $F(c,y)$ efficiently.

We modify the mass-concentration relation $c(M,z)$ from \cite{2008MNRAS.390L..64D}, $c_{\rm D}(M,z)$ to account for the baryonic feedback, following the treatment in HMx \citep{2020A&A...641A.130M}, i.e.,
\begin{equation}
    \frac{c(M,z)}{c_{\rm D}(M,z)} = 1+\epsilon_1+(\epsilon_2-\epsilon_1)\frac{f_{\rm bnd}(M)}{\Omega_{\rm b}/\Omega_m}=1+\epsilon_1+\frac{\epsilon_2-\epsilon_1}{1+(M_0/M)^\beta}~,
\label{eq:cM}
\end{equation}
where $\epsilon_1,\epsilon_2$ are free parameters and $\epsilon_1=\epsilon_2=0$ reduces to the unmodified case.

To eliminate the unphysical behaviour of the 1h-term at $k\rightarrow 0$, we adopt the treatment in the public halo model code \textsc{HMcode-2020} \citep[Eq.~17 of][]{2021MNRAS.502.1401M} for all 1h terms at low-$k$,
\begin{equation}
    P_{XY}^{\rm 1h}(k,z)\rightarrow P_{XY}^{\rm 1h}(k,z)\frac{(k/k_*)^4}{1+(k/k_*)^4}~,
\end{equation}
and take $k_*$ as the fitted functional form in their Table 2.

Our analysis assumes General Relativity (GR), flat background geometry, and the $w_0$-$w_a$CDM cosmology, \ie, a cold dark matter Universe with a time-varying dark energy component with its equation of state parameterised as $w(a) = w_0 + w_a(1-a)$ \citep{2001IJMPD..10..213C,2003PhRvL..90i1301L}.


\section{Inference details: Systematics, Covariances, Likelihood}
\label{sec:like}
Below we describe the implementation of survey specific details of our modelling, such as parameterisations of systematics, calculation of correlated statistical uncertainties (covariances), and other inference details. 

\subsection{Systematics}\label{ssec:sys}
Systematic uncertainties are parameterised through nuisance parameters, whose fiducial values and priors are summarised in Table~\ref{tab:params}. We closely follow the Roman forecasting papers of \cite{esk21,emk21} with the main difference that we use the halo model to capture uncertainties due to baryons instead of the simulation based principal component approach \citep[e.g.,][]{2019MNRAS.488.1652H,2021MNRAS.502.6010H}

\paragraph*{Photometric redshift uncertainties}
We parameterise the photometric redshift distributions of lens and source samples as 
\begin{equation}
    n_x(z_{\rm ph}) \equiv \frac{dN_x}{dz_{\rm ph}d\Omega} \propto z_{\rm ph}^2\exp[-(z_{\rm ph}/z_0)^\alpha]~,~~x\in\{{\rm lens,~source}\}
    \label{eq:nz-true}
\end{equation}
normalised by the effective number density $\bar{n}_x$. $N_x$ is the number counts of lens/source galaxies, $z_{\rm ph}$ is the photometric redshift, $\Omega$ is the solid angle. Each sample is further divided into 10 equally populated tomographic bins, with $i$-th bin distribution denoted by $n_x^i(z_{\rm ph})$.

We model photometric redshift uncertainties through Gaussian scatter $\sigma_{z,x}$ for lens and source sample each, and a shift parameter $\Delta_{z,x}^i$ for each redshift bin $i$ of the lens and source samples, such that the binned true redshift distribution $n_x^i(z)$ is related to the binned photometric redshift distribution $n_x^i(z_{\rm ph})$ by
\begin{equation}
    n_x^i(z) = \int_{z_{{\rm min},x}^i}^{z_{{\rm max},x}^i}\frac{dz_{\rm ph}\,n_x^i(z_{\rm ph})}{\sqrt{2\pi}\sigma_{z,x}(1+z_{\rm ph})}\exp\left[-\frac{(z-z_{\rm ph}-\Delta_{z,x}^i)^2}{2[\sigma_{z,x}(1+z_{\rm ph})]^2}\right]~.
\label{eq:binned-nz-true}
\end{equation}
In total, we have 22 photo-z parameters (10 shift parameters and 1 scatter parameter for lens and source sample each). These parameters are marginalised over using Gaussian priors. Details of the galaxy samples considered in this paper are further described in Sect.~\ref{sec:RomanCMB}.

\paragraph*{Linear galaxy bias}
We assume one linear bias parameter $b_{\rm g}^i$ per lens redshift bin. The fiducial values are assumed to follow the simple relation: $b_{\rm g}^i=1.3+0.1i$. The total of 10 linear bias parameters are independently marginalised over with a conservative flat prior $[0.4,3.0]$. We note that this model may be too simple for scales at $k_{\rm max}=$0.3$h/$Mpc, but since this is a forecasting paper this will not result in any parameter biases. The trade space of increased galaxy bias model complexity versus small-scale cosmological information gain is an important topic of future studies, especially in the context of CMB cross-correlations. 

\paragraph*{Multiplicative shear calibration}
We assume one parameter $m^i$ per source redshift bin, which affects $\kappa_{\rm g}$-$X$ ($X\in\lbrace \kappa_{\rm g}, \delta_{\rm g},\kappa_{\rm CMB},y \rbrace$) via
\begin{equation}
    q_{\kappa_{\rm g}}^i\rightarrow (1+m^i)q_{\kappa_{\rm g}}^i
\end{equation}
The total of 10 $m^i$ parameters are independently marginalised over with Gaussian priors.

\paragraph*{Intrinsic alignment (IA)}
We adopt the ``nonlinear linear alignment'' model \citep{2004PhRvD..70f3526H,2007NJPh....9..444B}, which considers only the ``linear'' response of the elliptical (red) galaxies' shapes to the tidal field sourced by the underlying ``nonlinear'' matter density field.
Our implementation follows \cite{2016MNRAS.456..207K} for cosmic shear, \cite{2017MNRAS.470.2100K} for galaxy-galaxy lensing, \cite{2022MNRAS.509.5721F} for $\kappa_{\rm g}$-$\kappa_{\rm CMB}$ power spectra, and extend it to $\kappa_{\rm g}$-$y$ power spectra. Using the notation in Sect.~\ref{ssec:probe-modeling}, we can encapsulate the effect as
\begin{equation}
    q_{\kappa_{\rm g}}^i\rightarrow q_{\kappa_{\rm g}}^i + q_I^i~,
\end{equation}
where
\begin{equation}
    q_I^i = -A(z)\frac{n_{\rm source}^i(z(\chi))}{\bar{n}_{\rm source}^i}\frac{dz}{d\chi}~.
\end{equation}
$A(z)$ is the IA amplitude at a given redshift $z$, computed by
\begin{equation}
    A(z)=\frac{C_1\rho_{\rm cr}}{G(z)}A_{\rm IA} \left(\frac{1+z}{1+z_0}\right)^{\eta_{\rm IA}}~,
\end{equation}
with pivot redshift $z_0=0.3$, $C_1\rho_{\rm cr}=0.0134$ derived from SuperCOSMOS observations \citep{2004PhRvD..70f3526H,2007NJPh....9..444B}. The fiducial values and priors for the nuisance parameters $A_{\rm IA},\eta_{\rm IA}$ are given in Table \ref{tab:params}. We neglect luminosity dependence and additional uncertainties in the luminosity function, which can be significant and are discussed in \cite{2016MNRAS.456..207K}.

Together with the multiplicative shear calibration, $q_{\kappa_{\rm g}}^i$ is altered as
\begin{equation}
    q_{\kappa_{\rm g}}^i(\chi)\rightarrow (1+m^i)[q_{\kappa_{\rm g}}^i(\chi)+q_I^i(\chi)]~.
\end{equation}

We do not consider higher-order tidal alignment, tidal torquing models \citep[see \eg,][]{2015JCAP...08..015B,2019PhRvD.100j3506B}, or more complicated IA modelling as a function of galaxy colour \citep{2019MNRAS.489.5453S}, or IA halo model \citep{2021MNRAS.501.2983F}. Similar to nonlinear galaxy bias models, the degrees of freedom that are opened up by these models may degrade the constraining power of the galaxy survey and enhance the importance of the information carried by secondary CMB effects.

\begin{table}
\footnotesize
\centering
\caption{\label{tab:params}The parameters characterising the Roman reference survey and CMB experiments, cosmology and systematics. The fiducial values are used for generating the simulated data vectors, and the priors are adopted in the sampling. Uniform priors are described by $U$[min, max], and Gaussian priors are described by $\mathcal{N}(\mu, \sigma^2)$.}
\begin{tabular}{l l l }
\hline\hline
Parameter & Fiducial & Prior \\  
\hline 
\multicolumn{3}{l}{\tbf{Survey}} \\
$\Omega_{\mathrm{s}} (\rm{Roman})$ & 2,000 deg$^2$ & -\\
$\Omega_{\mathrm{s}} (\rm{SO/S4})$ & 16,500 deg$^2$ & -\\
$n_{\mathrm{source}}$ & 51  arcmin$^{-2}$ & -\\
$n_{\mathrm{lens}}$ & 66  arcmin$^{-2}$& - \\
$\sigma_\epsilon $ &0.26/component& -\\
\hline 
\multicolumn{3}{l}{\tbf{Cosmology}} \\
$\om$ & 0.3156 & $U$[0.1, 0.6]   \\ 
$\sig$& 0.831  & $U$(0.6, 0.95]  \\ 
$\ns$ & 0.9645 & $U$[0.85, 1.06] \\
$\w$  & -1.0   & $U$[-2.0, 0.0]  \\
$\wa$ & 0.0    & $U$[-2.5, 2.5]  \\
$\omb$& 0.0492 & $U$[0.04, 0.055]\\
$h_0$ & 0.6727 & $U$[0.6, 0.76]  \\
\hline
\multicolumn{3}{l}{\tbf{Galaxy bias}} \\
$b^i$ ${\scriptstyle(i=1,\cdots,N_{\mathrm{tomo}})}$ & $1.3 + 0.1\times i$  & $U$[0.8, 3.0] \\
\hline
\multicolumn{3}{l}{\tbf{Photo-z}} \\
$\Delta_{z,\rm lens}^i$ ${\scriptstyle(i=1,\cdots,N_{\mathrm{tomo}})}$ & 0.0 & $\mathcal{N}(0, 0.001^2)$ \\
$\sigma_{z,\rm lens} $ & 0.01 & $\mathcal{N}(0.01, 0.002^2)$ \\
$\Delta_{z,\rm source}^i$ ${\scriptstyle(i=1,\cdots,N_{\mathrm{tomo}})}$ & 0.0 &$\mathcal{N}(0.0, 0.001^2)$ \\
$\sigma_{z,\rm source}$ &0.01 & $\mathcal{N}(0.01, 0.002^2)$ \\
\hline
\multicolumn{3}{l}{\tbf{Shear calibration}} \\
$m_i $ & 0.0 & $\mathcal{N}(0.0, 0.002^2)$\\
\hline
\multicolumn{2}{l}{\textbf{IA}} & \\
    $A_{\rm IA}$ & 0.5 & $U$[-5, 5]\\
    $\eta_{\rm IA}$ & 0.0 & $U$[-5, 5]\\
    \hline
\multicolumn{2}{l}{\textbf{Halo and Gas Parameters}} & \\
    $\Gamma$ & 1.17 & $U$[1.05, 1.35]\\
    $\beta$ & 0.6 & $U$[0.2, 1.0] \\
    $\log_{10}M_0$ & 14.0 & $U$[12.5, 15.0]\\
    $\alpha$ & 1.0 & $U$[0.5, 1.5]\\
    $\log_{10} T_w$ & 6.5 & $U$[6.0, 7.0] \\
    $\epsilon_1$ & 0.0 & $U$[-0.8, 0.8]\\
    $\epsilon_2$ & 0.0 & $U$[-0.8, 0.8]\\
    $f_{\rm H}$ & 0.752 & $U$[0.7, 0.8] \\
    $M_{\rm min}, M_{\rm max}$ & $10^6, 10^{17}M_\odot/h$ & fixed \\
    $A_*$ & 0.03 & fixed \\
    $\log_{10}M_*$ & 12.5 & fixed \\
    $\sigma_*$ & 1.2 & fixed \\
\hline\hline
\end{tabular}
\end{table}

\subsection{Covariances, Likelihood, Sampling}
\label{sec:covs}
The multi-probe data vector, denoted as $\D$, is computed at the fiducial cosmology and systematics parameter values (see Table \ref{tab:params}). The same parameters are assumed in the computation of the non-Gaussian covariance matrix, $\matC$. Given that this covariance matrix is calculated analytically, it is not an estimated quantity derived from either simulations or measured data. As a quantity that is free of estimator noise analytical covariance matrices can be inverted directly and do not require large amounts of realisations for the inverse to be precise \citep[see e.g.,][for details on the number of realizations and alternative ideas]{2013MNRAS.432.1928T,2013PhRvD..88f3537D,2018MNRAS.473.4150F}. 

We sample the joint parameter space of cosmological $\pco$ and nuisance parameters $\pnu$, the latter describing our systematic uncertainties, and parameterise the joint likelihood as a multivariate Gaussian 
\begin{equation}
\label{eq:like}
\like (\D| \pco, \pnu) = N \, \times \, \exp \biggl( -\frac{1}{2} \underbrace{\left[ (\D -\M)^t \, \matC^{-1} \, (\D-\M) \right]}_{\chi^2(\pco, \pnu)}  \biggr) \,.
\end{equation}
The model vector $\M$ is a function of cosmology and nuisance parameters, i.e. $\M=\M(\pco, \pnu)$ and the normalisation constant $N=(2 \pi)^{-\frac{n}{2}} |\matC|^{-\frac{1}{2}}$ can be ignored under the assumption that the covariance is constant while the MCMC steps through the parameter space. 

We calculate the covariance of two angular power spectra as the sum of the Gaussian covariance, $\matC^{\rm G}$, and non-Gaussian covariance in the absence of survey window effects $\matC^{\rm cNG}$, and the super-sample covariance, $\matC^{\rm SSC}$, which describes the uncertainty induced by large-scale density modes outside the survey window
\begin{equation}
    \matC= \matC^{\rm G}+ \matC^{\rm cNG} + \matC^{\rm SSC} \,.
\end{equation}

We assume a multivariate Gaussian likelihood as described in Eq. (\ref{eq:like}) with a constant covariance matrix computed at the fiducial parameters. 
    
We sample the parameter space using \texttt{emcee\footnote{\url{https://emcee.readthedocs.io/en/stable/}}} \citep{2013PASP..125..306F}, which is based on the affine-invariant sampler of  \cite{2010CAMCS...5...65G}, and which can be parallelised with either MPI or shared memory multiprocessing. 

\section{Roman reference survey design - CMB synergies}
\label{sec:RomanCMB}

In this section we present results for the joint analyses of the 2000 deg$^2$ Roman reference survey assuming overlap with Rubin's LSST, and two CMB experiments, namely CMB-S4 (S4) and Simons Observatory (SO). 

\paragraph*{Roman HLWAS reference survey}
The Roman HLWAS reference survey design ensures excellent systematics control via space-quality imaging and photometry across 4 bands: F106, F129, F158, F184 in the near-infrared\footnote{\url{https://roman.gsfc.nasa.gov/science/WFI_technical.html}}. This Roman coverage is complemented by 6 bands in the optical from LSST at similar depth. The reference survey design covers 2000 deg$^2$ and takes $\approx$1 year.

We assume the same redshift distributions as in \cite{emk21}, which are obtained by applying results from the Roman exposure time calculator \citep{hgk12} to the CANDELS dataset \citep[see][for further details]{2019ApJ...877..117H}. The exact criteria are: 

\begin{itemize}
\item \tbf{Clustering or lens sample:} The standard lens sample is comprised of galaxies with two main selection criteria: $S/N$>10 in each of the 4 Roman bands and $S/N$>5 in each LSST band, except for the $u$-band. We note that nevertheless more than 50\% of our galaxies have $S/N$>5 in the $u$-band.
\item \tbf{Weak lensing or source sample:} Starting from the clustering sample the weak lensing sample  requires $S/N$>18 ($J+H$ band combined, matched filter), resolution factor R>0.4, and ellipticity dispersion $\sigma_\epsilon<0.2$.
\end{itemize}

\paragraph*{CMB-S4}
The CMB-S4 experiment \cite{2016arXiv161002743A} is a joint DOE/NSF project that consists of both large and small ground-based telescopes that are located in the Atacama desert in Chile and at the South Pole. It will cover a large fraction of the sky in 11 observing bands to remove dust and synchrotron foregrounds. With its high resolution (beam FWHM close to 1 arcmin) and sensitivity (white noise level of the order $\sim$ 1$\mu$K$\cdot$arcmin) its noise CMB lensing reconstruction noise and tSZ signal to noise is significantly improved over SO. In terms of survey area we assume the same 16,500 deg$^2$ area coverage as for SO. Differences in constraining power are hence due to the increased sensitivity. 

The CMB-S4 noise curves are generated with a very similar procedure as the ones for SO, but based on the survey performance expectations publicly available on the CMB-S4 wiki\footnote{\url{https://cmb-s4.uchicago.edu/wiki/index.php/Survey_Performance_Expectations}}.

\paragraph*{Simons Observatory}
The Simons Observatory (SO, \citealp{2018SPIE10708E..04G,2019JCAP...02..056A}) is a CMB experiment under construction in the Atacama desert in Chile, at an altitude of 5,200~m. It is designed to observe the microwave sky in six frequency bands centred around 30, 40, 90, 150, 230, and 290~GHz, in order to separate the CMB from Galactic and extragalactic foregrounds. The observatory will include one 6~m large-aperture telescope (LAT, \citealp{2021RNAAS...5..100X,2018SPIE10700E..41P}) and three small-aperture 0.5~m telescopes (SATs, \citealp{2020JLTP..200..461A}).
The LAT will produce temperature and polarisation maps of the CMB with $\sim$arcminute resolution over 40\% of the sky, with a sensitivity of $\sim$6 $\mu$K$\cdot$arcmin when combining 90 and 150~GHz bands.
These wide deep maps will be the key input to measure CMB lensing with SO.

The noise models for SO are based on component separated maps for the ``goal'' sensitivity (\texttt{SENS-2}) including the expected level of residual foregrounds and atmospheric noise. The noise for the tSZ map\footnote{\url{https://github.com/simonsobs/so_noise_models/blob/master/LAT_comp_sep_noise/v3.1.0/SO_LAT_Nell_T_atmv1_goal_fsky0p4_ILC_tSZ.txt}} is obtained from the standard ``Internal Linear Combination'' (ILC, \texttt{deproj0}) without any additional deprojection on $40\%$ of the sky. The CMB lensing map\footnote{\url{https://github.com/simonsobs/so_noise_models/blob/master/LAT_lensing_noise/lensing_v3_1_1/nlkk_v3_1_0_deproj0_SENS2_fsky0p4_it_lT30-3000_lP30-5000.dat}} is reconstructed with the same sensitivity and area as the tSZ map, and with the minimum variance ILC including both temperature and polarisation and no other deprojection (\texttt{deproj0}). A small improvement expected from iterative lensing reconstruction is also included in the lensing noise. More details on how the noise curves are generated can be found in Sect.~2 of \cite{2019JCAP...02..056A}.

Finally, we note that a recent investment in SO will increase the detector count and lower the noise, in an upgrade known as ``Advanced Simons Observatory'' (ASO). Little public  information is available at the time of writing, and therefore our forecasts adopt the nominal configuration. Further gains beyond what is presented in this work can be expected with ASO.

\subsection{Multi-probe data vectors}
\label{sec:data}
Our 10\x2 data vector is composed of 10 different auto- and cross power spectra, composed of all permutations (ignoring symmetries) of the underlying fields, i.e. $X\in\lbrace \kappa_{\rm g}, \delta_{\rm g},\kappa_{\rm CMB},y \rbrace$. We also consider subsets of 10\x2 such as a 6\x2 data vector, which leaves out the $y-$field, and a 3\x2 data vector that leaves out both the $y$ and the $\kappa_{\rm CMB}$ fields.

We adopt 25 logarithmically spaced Fourier mode bins ranging from $\ell_{\rm min}=20$ to $\ell_{\rm max}=8000$ for all two-point functions, and adopt the following set of $\ell$-cut:
\begin{itemize}
    \item \textbf{Galaxy clustering $\delta_{\rm g}$} scale cuts are driven by the modelling inaccuracy of nonlinear galaxy biasing. We adopt $\ell^i_{\rm max}=k_{\rm max}\,\chi(\bar{z}^i)$, where $k_{\rm max}=0.3\,h/$Mpc and $\bar{z}^i$ is the mean redshift of the lens bin $i$.
    \item \textbf{Weak lensing $\kappa_{\rm g}$} scale cuts are driven by model misspecifications for the impact of baryonic processes on the nonlinear matter power spectrum as well as gravitational nonlinearity.
    
    We assume $\ell_{\rm max}=4000$ for all tomography bins. We note that baryonic effects on the shear power spectrum will be significant for both of these scale cuts, and Stage-IV analyses will likely require modelling beyond the modified halo mass-concentration relation (Eq.~\ref{eq:cM}) implemented here to reach these scale cuts, these different $\ell_{\rm max}$ choices will illustrate the interplay of scale cuts and parameter degeneracies. 
    \item \textbf{CMB lensing $\kappa_{\rm CMB}$} measurements are challenging at high lensing $\ell$ due to large reconstruction noise and uncertainties from foregrounds; similarly, various systematic errors and mean field uncertainties can limit the reconstruction of low-$\ell$ lensing modes \citep{2021MNRAS.500.2250D}. We adopt the scale cuts $\ell_{\rm min}=100$, $\ell_{\rm max}=3000$ for the CMB lensing map.\footnote{While the interpretation of CMB lensing (cross-) spectra is of course also limited by the same high-$\ell$ processes that set the galaxy weak lensing scale cuts, the fractional impact of these processes on the power spectrum decreases with redshift. Hence $\ell_{\rm max}=3000$ for $\kappa_{\rm g}$ would correspond to a less restrictive scale cut for $\kappa_{\rm CMB}$.}
    \item \textbf{tSZ} $y$ measurements are limited by atmospheric noise at low-$\ell$ (in the case of SO) and component separation at high-$\ell$ \citep{2019JCAP...02..056A}. We adopt the somewhat optimistic scale cuts $\ell_{\rm min}=80$, $\ell_{\rm max}=8000$, noting in particular that contamination from the Cosmic Infrared Background (CIB) will need to be modelled and (likely) marginalised over.
\end{itemize}
For cross-power spectra, we adopt the more restrictive scale cut combination of the two fields. We further exclude $\delta_{\rm g}$-$\kappa_{\rm g}$ combinations without cosmological signal, \ie, with the lens tomography bin at higher redshift than the source tomography bin.

\subsection{Results}
\label{sec:R2kresults}

\begin{figure}
\includegraphics[width=8.8cm]{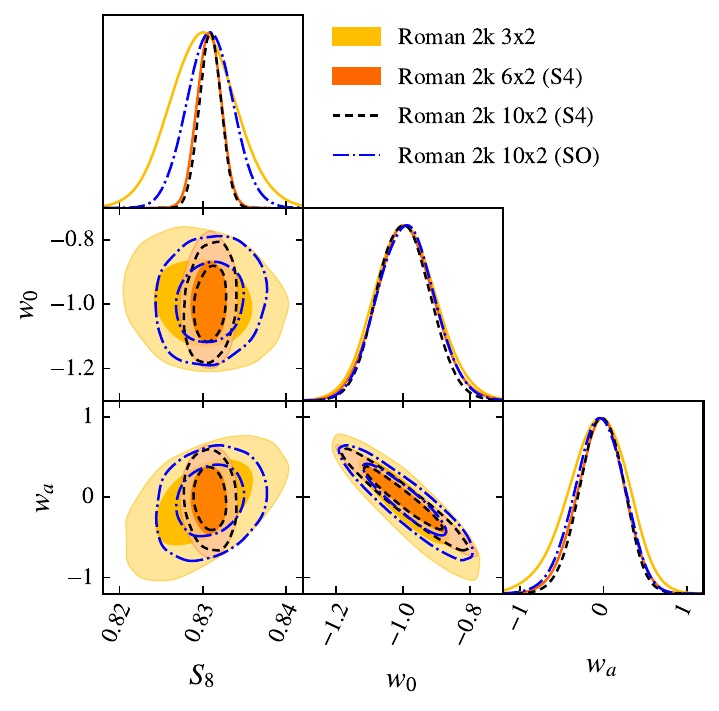}
\caption{This figure shows the increase in constraining power as a function of increasing the data vector from 3\x2 (yellow) to 6\x2 (orange) to 10\x2 (black dashed) for the Roman reference (Roman 2k) survey in combination with CMB-S4. For comparison we also show the 10\x2 constraints when replacing S4 with Simons Observatory (blue, dashdotted).}
\label{fi:3610}
\end{figure}

\begin{table}
\footnotesize
    \centering
    \caption{\label{tab:snr1} The signal-to-noise ratios (S/N) and dark energy figure of merit values for the Roman 2k + S4/SO scenarios studied. The brackets show the \%-value increase over the Roman 2k scenario.}
    \begin{tabular}{r c c c}
    \hline\hline
    Probe & Roman 2k+ SO & Roman 2k + S4  \\
    \hline
    \multicolumn{3}{c}{\tbf{SNR }} \\ 
    \hline
    3\x2pt & 530 & 530  \\
    6\x2pt & 551 (4\%) & 599  (13\%)\\
    10\x2pt& 635 (20\%) & 786  (48\%) \\
    \hline
     \multicolumn{3}{c}{\tbf{DETF-FOM}} \\
    \hline
    3\x2pt & 64 & 64   \\
    6\x2pt & 75 (17\%) & 143 (123\%)  \\    
    10\x2pt& 92 (48\%) & 172 (168\%) \\
    \hline\hline

    \end{tabular}
\end{table}

As summarized in Table \ref{tab:snr1} and illustrated in Fig. \ref{fi:3610}, we find a significant increase in constraining power when adding CMB lensing information from S4 to the 3\x2pt data vector, consistent with previous forecast studies \citep[\eg,][]{2022MNRAS.509.5721F,2022MNRAS.512.5311W,2024MNRAS.527.9581F}. 

We find a modest gain in SNR (13\% for 6\x2 and 48\% for 10\x2, respectively) and a significant gain in the dark energy FOM (123\% for 6\x2 and 168\% for 10\x2, respectively). The bulk of this information gain can be attributed to the auto-correlation of CMB lensing and $y$-maps, and the cross-correlation between $\kappa_\mr{CMB}$ and $y$. The smaller Roman survey area determines the fraction of the sky over which cross-correlations between galaxy maps (density and shear) with CMB maps $\kappa_\mr{CMB}$ and $y$-maps can be measured. Given the $\sim$8 times smaller area of Roman 2k, the corresponding information gain from these cross-correlations is expected to less important.

Similar to \citep{2024MNRAS.527.9581F} we find that adding tSZ to 6\x2pt causes a less prominent increase in constraining power compared to adding CMB lensing to 3\x2 due to the additional parameter degeneracies introduced by the tSZ model \citep{2024MNRAS.527.9581F}.

We find similar effects but less pronounced improvements on SNR and dark energy FOM when using Simons Observatory, which is a consequence of the comparably larger noise-levels of the CMB observables compared to CMB-S4.

\section{Roman wide survey concepts}
\label{sec:wide}
Of particular interest for synergies with CMB experiments are Roman wide survey concepts. Future CMB experiments like the here considered SO/S4 surveys cover 16,500 deg$^2$, which would enable high precision cross-correlation science if the same area were observed with Roman as well.

A wider survey can be achieved by increasing the survey duration using the reference design, decreasing the depth in each band, or changing from the nominal 4-band survey to a single-band survey in either the H-band or the even broader W-band (F146). In this scenario multi-band photometry is provided by the  LSST dataset and the single band of Roman data would contribute space-based shape measurements, high-resolution data to inform blending, and observer-frame NIR photometry and morphology information.

Our starting point for exploring Roman wide survey concepts is the analysis in \cite{esk21}, where the authors explore a deep 18,000 deg$^2$ single-band Roman survey design (W-band, F146) that overlaps with the 6 LSST bands and takes 1.5 years to complete. Adopting their selection criteria for suitable galaxies results in a lower number density in the lens and source sample (50 and 43, respectively) compared to the 4-band reference design (66 and 51, respectively). We also follow their analysis in assuming an increased photo-z uncertainty (cf. Table \ref{tab:params_wide} and Table \ref{tab:params}) due to the fact that the wavelength range previously covered in 4 bands is now covered in the W-band only.

\begin{figure}
\includegraphics[width=8.5cm]{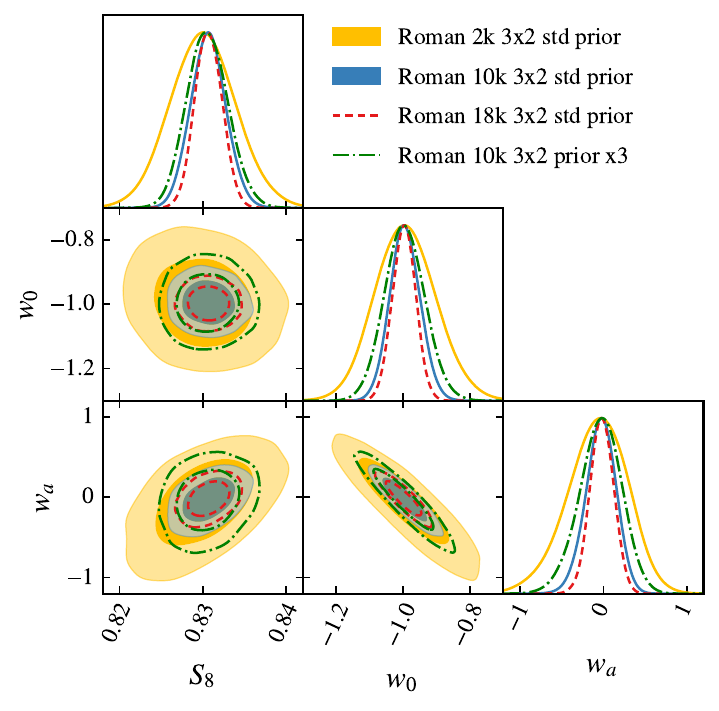}
\caption{This figure shows the increase in constraining power as a function of increasing the area of the Roman survey from 2000 deg$^2$ to 10,000 deg$^2$ to 18,000 deg$^2$ for 3\x2.}
\label{fi:area}
\end{figure}

Comparing the Roman 2k and Roman 18k scenarios we find an increase in the dark energy figure of merit by a factor of 6.8 (cf. solid yellow and red dashed contour in Fig. \ref{fi:area}), which is slightly larger compared to the factor of 5.5 found in \cite{esk21}. This discrepancy can be explained by the different approaches to marginalize over baryonic uncertainties and the related different parameter spaces used in the analysis. Directionally, the findings of both analyses are similar and demonstrate that wide scenarios are an interesting concept to explore further given that the increase in survey area seems to outweigh the decrease in number density of galaxies and increase in systematic uncertainties. 

In addition to the Roman 18k scenario we consider a reduced survey footprint of 10,000 deg$^2$ with otherwise the exact same parameters. The Roman 10k scenario shows a factor of 4.2 improvement over the Roman 2k FoM. This is, of course, less than the Roman 18k FoM gain, but it would reduce the survey time from 1.5 years to 10 months.

\begin{table}
\footnotesize
\centering
\caption{\label{tab:params_wide} Parameters for the  Roman wide scenarios that differ from the values for Roman reference survey presented in Tab.~\ref{tab:params}.}
\begin{tabular}{l l l }
\hline\hline
Parameter & Fiducial & Prior \\  
\hline 
\multicolumn{3}{l}{\tbf{Survey}} \\
$\Omega_{\mathrm{s}} (\textrm{Roman; deg$^2$})$ & 10,000 or 18,000 & - \\
$n_{\mathrm{source}}$ & 43 arcmin$^{-2}$ & -\\
$n_{\mathrm{lens}}$ & 50 arcmin$^{-2}$ & - \\
\hline 
\multicolumn{3}{c}{\tbf{Standard systematic uncertainties}} \\
\hline 
\multicolumn{3}{l}{\tbf{Photo-z}} \\
$\sigma_{z,\rm lens} $ & 0.02 & $\mathcal{N}(0.02, 0.002^2)$ \\
$\sigma_{z,\rm source}$ &0.02 & $\mathcal{N}(0.02, 0.002^2)$ \\
\hline 
\multicolumn{3}{c}{\tbf{Strong systematic uncertainties }} \\
\hline 
\multicolumn{3}{l}{\tbf{Photo-z}} \\
$\Delta_{z,\rm lens}^i$ ${\scriptstyle(i=1,\cdots,N_{\mathrm{tomo}})}$ & 0.0 & $\mathcal{N}(0.0, 0.003^2)$ \\
$\sigma_{z,\rm lens} $ & 0.02 & $\mathcal{N}(0.02, 0.006^2)$ \\
$\Delta_{z,\rm source}^i$ ${\scriptstyle(i=1,\cdots,N_{\mathrm{tomo}})}$ & 0.0 &$\mathcal{N}(0.0, 0.003^2)$ \\
$\sigma_{z,\rm source}$ &0.02 & $\mathcal{N}(0.02, 0.006^2)$ \\
\multicolumn{3}{l}{\tbf{Shear calibration}} \\
$m_i $ & 0.0 & $\mathcal{N}(0.0, 0.006^2)$\\
\hline\hline
\end{tabular}
\end{table}

\subsection{Synergies with CMB-S4 surveys}

\begin{figure}
\includegraphics[width=8.5cm]{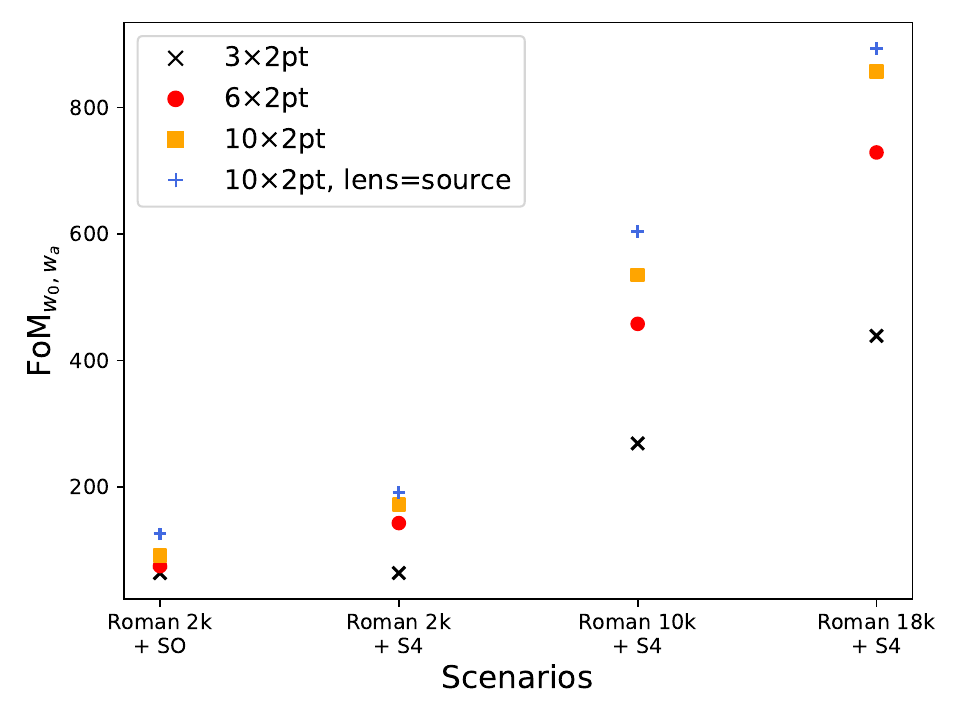}
\caption{The dark energy Figure-of-Merit (FoM) of 3\x2pt, 6\x2pt, and 10\x2pt analyses for different survey designs of the Roman Space Telescope. We find significant increases in constraining power when increasing the Roman survey area to 10k and 18k deg$^2$, respectively, for all 3 types of analyses. Interestingly, we also find that the one galaxy sample analyses (blue '+') generally have a higher FOMs, depsite the fact that this setup results in fewer galaxies and a shallower redshift distribution for the clustering sample. }
\label{fi:fom}
\end{figure}

As we can see from Fig. \ref{fi:fom} and Table \ref{tab:snr}, including CMB lensing and tSZ information, i.e. when going from 3\x2 to 10\x2, shows roughly a factor of 2 improvement for both wide scenarios. This is slightly less compared to the Roman 2k case, where we saw an almost 3-fold increase in the FoM. Given that the S4 survey area in all cases is 16,500 deg$^2$ this trend is expected. Comparing 3\x2 and 6\x2 we find a FoM gain that corresponds to approximately two-thirds of the gain we see for 10\x2 in both wide survey cases. 

We also explore the idea of using the source sample as the lens sample (``lens=source''), i.e., using the same galaxies for both clustering and lensing and in all cross-correlations with CMB probes. This idea has been discussed in \cite{2020JCAP...12..001S} and investigated in \cite{2022MNRAS.509.5721F} in the context of 6\x2pt analysis. Besides the obvious advantage of eliminating half of the photo-z nuisance parameters, simulated analyses have shown that it can lead to significant improvement in photo-z uncertainties and tighter constraints of some cosmological parameter combinations.

We find that the ``lens=source'' setup is outperforming the case of a lens sample that includes all galaxies suitable for clustering. While the latter has a higher galaxy number density and a redshift distribution that extends to higher redshift, the fact that 1 sample allows for improved self-calibration wins (see Fig. \ref{fi:fom} and Table \ref{tab:snr}). This self-calibration of photo-z uncertainties but also of baryonic physics uncertainties through tSZ and CMB lensing cross-correlations, translates into significant gains in SNR ($\sim 20\%$) but also into gains for the dark energy figure of merit (5-10\%), despite the reduction of the galaxy sample size.

We note the benefits of overlapping survey area of galaxy and CMB surveys that allow the 6\x2 and 10\x2 analyses self-calibrate galaxy bias and other systematics parameters in high-redshift bins, through the cross-correlations between CMB lensing and galaxy fields. These high redshift observational systematics are notoriously difficulty to model/mitigate with galaxy surveys alone. 

\begin{table}
\footnotesize
    \centering
    \caption{\label{tab:snr}The signal-to-noise ratios (S/N) and dark energy FoMs of the different combined probes scenarios for the Roman 10k and 18k surveys.}
    \begin{tabular}{r c c}
    \hline
     \tbf{Probe} &  \tbf{Roman 10k + S4} & \tbf{Roman 18k + S4}  \\
    \hline
   \multicolumn{3}{c}{\tbf{SNR }} \\ 
    \hline
    3\x2pt & 1118  & 1508  \\
    6\x2pt & 1161  & 1551   \\
    10\x2pt& 1313  & 1715   \\
   (lens=source) 10\x2pt& 1642 & 2167 \\\hline
    \multicolumn{3}{c}{\tbf{DETF-FOM}} \\
    \hline
    3\x2pt & 269  & 439   \\
    6\x2pt & 458  & 729   \\
    10\x2pt& 535  & 857   \\
    (lens=source) 10\x2pt & 604 & 893    \\
    \hline
    \multicolumn{3}{c}{\tbf{DETF-FOM 3x systematic uncertainties}} \\
    \hline
    10\x2pt& 178 & -  \\
    \hline\hline
    \end{tabular}
\end{table}

\subsection{Systematics study of Roman wide scenarios}
\label{sec:wide-sys}
The Roman wide scenarios boost the statistical constraining power significantly, and it requires careful study of systematics control. Of particular concern are photo-$z$ uncertainties, due to the limited information on the spectral energy distributions (SEDs) at $>1\,\mu$m, and shape measurement uncertainties, due to the reduced dithering and the wavelength dependence of the PSF.

One goal of the Roman Space Telescope survey design is to understand how far Roman can push its wide survey idea before being endangered by systematics. The interplay between statistical and systematic error contributions is subtle due to the high dimensionality of the data vector space. In a one-dimensional data vector space, only one parameter can be constrained, and it has total uncertainty $\sigma = \sqrt{\sigma_{\rm stat}^2 + \sigma_{\rm sys}^2}$, with $\sigma_{\rm stat}^2\propto 1/N$ (where $N$ is sample size). This leads to the familiar notion that once the sample size is large enough to make $\sigma_{\rm stat}\ll\sigma_{\rm sys}$, then one is ``systematics-limited'' in the sense that the benefits from larger sample size have saturated ($\sigma\approx\sigma_{\rm sys}$). However, in a high-dimensional parameter space with many systematics, the systematic contribution to the covariance matrix can dominate in some directions in the parameter space while the statistical contribution dominates in others; and the cosmological parameters of interest are almost always in diagonal directions.\footnote{In formal linear algebra language, this statement means that $\partial{\boldsymbol D}/\partial p_\alpha$ is not a right eigenvector of ${\bf C}_{\rm sys}{\bf C}_{\rm stat}^{-1}$, except in accidental special cases.} When this happens, it can both be true that the systematic errors are ``dominant'' in the sense that the parameter constraints get better when the systematics priors are improved; but increasing the sample size (here: survey area) also leads to continued improvement as some systematics directions are self-calibrated. Thus it is possible for the systematic errors to be a major contribution to the error budget, but also not represent a ``floor:'' the optimal survey might still have larger area. Of course, it is still necessary to ensure that the systematics parameterization is flexible enough to accommodate the likely real scenarios (even if the parameters themselves are not known a priori).

While the wide survey scenarios described above take reduced galaxy number density and increased photo-z uncertainties into account, the exact increase in systematic uncertainties needs further study (including from ongoing work within the Roman groups). Below we consider several specific cases of increased systematics and how they would impact Roman results. 

\paragraph*{Impact of increased observational uncertainties} The green dot-dashed contours in Fig. \ref{fi:area} correspond to the Roman 10k scenario (blue solid) but assume increased uncertainties in observational systematics (both shear calibration and photo-z uncertainties). The exact parameter values are shown in Table \ref{tab:params_wide}, which correspond to three times the fiducial values.
 
The increased systematics reduce the Roman 10k 3\x2 FoM (178) by ~30\% compared to the 10k 3\x2 FoM with standard systematics (269), however this FoM is still almost a factor of 3 larger compared to the Roman 2k 3\x2 FoM (64). This indicates that increasing the Roman survey area by a factor of 5 compared to the reference design can be beneficial even if the change in survey strategy causes an increased systematics budget.

\paragraph*{Impact of imperfect astrophysical modeling}
In addition to the observational systematics, most prominently shear calibration and photo-z uncertainties, astrophysical uncertainties become  important to control when increasing the statistical power. As a prominent example we consider halo model uncertainties, more specifically we use two alternative mathematical descriptions for the halo-mass concentration relation \citep{2008MNRAS.390L..64D, ludlow16} to generate data vectors and subsequently analyse them with our standard likelihood model.

\begin{figure}
\includegraphics[width=8.1cm]{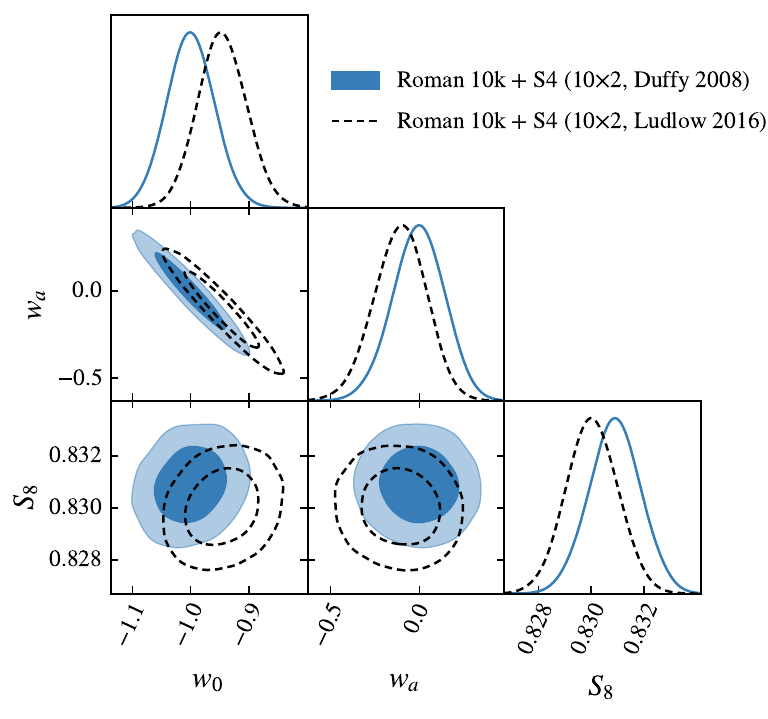}
\caption{The figure shows using our fiducial model to fit the data vector produced with $c(M)$ in Ludlow16 instead of Duffy08 will lead to significant bias in cosmology in the Roman 10k \x CMB-S4 10\x2pt analysis, assuming $w_0w_a$CDM cosmology.}
\label{fi:ludlow}
\end{figure}

Figure \ref{fi:ludlow} shows the bias in parameter space as a consequence of our different modelling assumptions in the data vector compared to the model pipeline for a Roman 10k 10\x2 analysis. We find biases significantly less than 1 $\sigma$ in the most relevant parameters, i.e. $w_0$, $w_a$, $S_8$. 

\section{Discussion of Roman survey design scenarios - a two-tier survey idea}
\label{sec:wide-designs}
As discussed, the reference design for the HLWAS covers 2000 deg$^2$ in 4 bands + the grism. A discussion of the grism survey trade space is outside the scope of this paper, but should be included in future studies.

Even when considering Roman science return independent of CMB survey synergies, there is a strong benefit in statistical errors to going wider. Such designs however run into the basic challenge that with an appropriate dithering pattern to achieve full sampling in 4 bands and with the read noise and slew-settle overheads of an infrared focal plane on a large observatory. Faster surveys become inefficient since the sensitivity degrades much faster than the square root of survey speed.

The most promising way to break free of this limitation is a two-tier HLWAS: splitting the time between a ``medium'' tier, which resembles the reference survey design but with a reduced area, and a ``wide'' tier in a single filter. The highly redundant medium tier would be used as an ``anchor'' for calibrating shear and photo-$z$ systematics in the wide tier. The following considerations assume that the survey time for any design is fixed to the survey time of the 2000 deg$^2$ reference survey.

Notionally, one might consider an imaging program for the medium tier with $\sim 1000$ deg$^2$ in 4 bands (or $\sim 800$ deg$^2$ in 5 bands if K band is added).  

For the wide single-filter tier a variety of choices exists and we show a comparison of the survey yields for the reference survey and three additional options in Table~\ref{tab:depths} and discuss them below:

\paragraph*{Option 1 (8000 deg$^2$/yr, H-band):} A wide layer done in the H band with the 2-pass reference survey strategy. The H band would be chosen because it avoids the thermal background of F184, but has better sampling properties than Y and J. This choice is relatively conservative, in that it recovers full sampling using IMCOM \citep{2024MNRAS.528.2533H} and enables \"ubercalibration using the cross-linked survey strategy. It can use the same data processing tools with similar settings as the medium-tier survey, and would serve as 1 of the 4 bands if the decision were made to increase the area of the medium tier in an extended mission.
There will be enhanced photo-$z$ scatter due to the 1.06--1.38 $\mu$m gap in photometric coverage; the source redshift distributions in the wide layer would have to be calibrated from the medium layer using the mapping from the full color space to the lower-dimensionality $ugrizyH$. The principal disadvantage is that the 1-band survey speed is only $\sim 4\times$ faster than the medium tier 4-band survey speed.
\paragraph*{Option 2 (13,000 deg$^2$/yr, H-band):} A variant is to do the wide layer in the H band with reduced dithers (one full pass, and only a sparse second pass for cross-linking) to save time. The disadvantage is that full sampling is not recovered, so in addition to calibrating the effect on photo-$z$s of having only some of the bands, one would also have to calibrate the undersampling effects. Since the currently planned Roman shear pipeline is based on shear calibration techniques that use full sampling in an essential way \citep{2017arXiv170202600H, 2024MNRAS.52710388L}, it would be necessary to develop an additional module to transfer the shear calibration from the Medium to the Wide layer.
\paragraph*{Option 3 (13,000 deg$^2$/yr, W-band):} The fastest weak lensing survey possible with Roman is to use the W filter with shorter exposures. This opens the exciting possibility of observing the full extragalactic LSST footprint. However, it also poses the greatest challenge for systematics: in addition to needing to calibrate $ugrizyW$ photo-$z$s and undersampling effects from the 4-band layer, we would need to calibrate the chromatic astrometric and PSF effects over the 0.93--2.00 $\mu$m bandpass \citep{cav10, csl18, erh19,2021MNRAS.501.2044T, 2024MNRAS.528.6680Y}.

\begin{table}
\caption{\label{tab:depths} The forecast survey yields of the reference survey and the various wide options. Shape measurement cuts are defined with matched filter S/N$>18$; resolution factor $>0.4$; and ellipticity measurement error per component $\sigma_e<0.2$ (in the \citet{2002AJ....123..583B} convention). Combinations such as ``J+H'' denote a synthetic filter made by combining, e.g., the J and H band images. The zodiacal light brightness of 1.5$\times$pole is $\sim$70th percentile in the reference survey design; the case of $2.5\times$pole is appropriate for regions near the Ecliptic in the wide survey options.}
\begin{tabular}{cccc}
\hline\hline
Case & band & Zodi brightness & 5$\sigma$ pt src depth \\
 & & [pole=1] & [mag AB] \\
\hline
Reference Survey
 & F184 & 1.5 & 26.1 \\
(2000 deg$^2$/yr) & H & 1.5 &  26.7 \\
 & J & 1.5 & 26.8 \\
& Y & 1.5 &  26.7 \\
 & J+H & 1.5 & 27.1 \\
\hline
Wide tier, Option 1
 & H & 1.5 &  26.7 \\
(8000 deg$^2$/yr) & H & 2.5 & 26.5 \\
\hline
Wide tier, Option 2
 & H & 1.5 &  26.4 \\
(13000 deg$^2$/yr) & H & 2.5 &  26.2 \\
\hline
Wide tier, Option 3
 & W & 1.5 &  27.1 \\
(13000 deg$^2$/yr) & W & 2.5  & 26.9 \\
\hline\hline
\end{tabular}
\end{table}

\paragraph*{Footprints:} The sky area available for Roman + Rubin synergistic surveys could in principle be expanded to include almost half the sky. However, although the Universe as a whole is isotropic, observing conditions make some of these regions more valuable than others. The ``best'' area for the optical data will likely be the regions that pass near zenith at the Rubin site (to minimize atmospheric dispersion and extinction effects) and have low Galactic dust column (to minimize reddening). There are 7010 deg$^2$ of sky that pass within $30^\circ$ of zenith at Rubin and have $E(B-V)_{\rm SFD}<0.05$ mag.\footnote{This area varies continuously with the choice of boundaries, e.g., it is 11050 deg$^2$ at $E(B-V)_{\rm SFD}<0.1$ mag.} This region is split into 5780 deg$^2$ (South Galactic Cap) + 1230 deg$^2$ (North Galactic Cap).

We note that of this ideal region 5300 deg$^2$ are $>15^\circ$ from the Ecliptic. These regions farther from the Ecliptic have lower zodiacal background available from space, but they are also farther south and hence less accessible to Northern Hemisphere facilities. Because read noise is significant in the fastest Roman survey modes, the benefit of choosing high ecliptic latitude fields is not as strong as for Euclid/VIS: for example, in $5\times 140$ s exposures in Roman H-band, we find the weak lensing source density $n_{\rm eff}$ varies from 38 arcmin$^{-2}$ at the Ecliptic poles to 30 arcmin$^{-2}$ in the Ecliptic at quadrature (only a 23\% loss).

\section{Conclusions}
\label{sec:conc}
This paper builds on the 10\x2 analysis that includes weak lensing, galaxy clustering, CMB lensing, tSZ auto- and cross-correlations as described in \cite{2024MNRAS.527.9581F}. Here, we focus on the Roman Space Telescope and explore synergies with CMB experiments, CMB-S4 and Simons Observatory. We note that for all Roman scenarios we assume overlap with Rubin Observatory's multi-band photometric observations.

When combining the 2000 deg$^2$ Roman reference survey with SO we find moderate improvement in the dark energy figure-of-merit, even though CMB lensing and tSZ are computed over a much larger survey area (16,500 deg$^2$). This changes when combining Roman 2k with S4 over 16,500 deg$^2$, which results in more than doubling the DETF-FOM.

The more interesting scenarios unfold when considering Roman survey strategies that encompass 10,000 deg$^2$ and 18,000 deg$^2$, respectively. While these Roman wide scenarios assume lower galaxy density, shallower redshift distributions, and increased systematic uncertainties compared to Roman 2k, they all demonstrate significant gains in cosmological information. Even in the case where we have tripled the range of uncertainties in photo-z and shear calibration priors, we find a significant increase in the dark energy FOM, when comparing Roman 10k to Roman 2k for the case of a standard 3\x2 analysis. 

The principal risk is the complexity of the systematics model associated with an increased survey area: inhomogeneous sampling or noise effects, large-scale zero-point variations, and (in W band) hidden correlations of the shear calibration error with other source galaxy properties may lead to additional systematics parameters that have to be inferred from the data vector beyond those considered here. If these issues turn out to be severe enough, a survey analogous to Option 1 might be chosen. Nevertheless, the potential dramatic improvements from reaching $\ge 10,000$ deg$^2$ are large and Options 2 and 3 should be studied, even if only for an extended mission. 

This statement is even more true in light of synergies with S4. The joint coverage of the Roman survey area with CMB information enables multiple routes for self-calibration of observational and astrophysical systematics. Thermal SZ (and kinetic SZ) are emerging as powerful probes to self-calibrate baryonic physics \citep[e.g.][]{2021PhRvD.103f3514A, 2022PhRvD.105l3525G, 2023MNRAS.525.1779P, 2024arXiv240406098B}. CMB lensing information and cross-correlations with the galaxy density and shear field help calibrate some combination of galaxy bias, photo-z, and shear measurement uncertainties at higher redshifts that are otherwise difficult to control. The fact that this information would be available over 16,500 deg$^2$ would allow us to consider Roman wide survey ideas with limited risk.

We subsequently map out multiple Roman wide survey implementations that rely on a two-tier concept: 1) a medium tier that relies on the 4-band reference survey design but over a $\sim$ 1000 deg$^2$ area and 2) a wide tier that covers a larger fraction of the sky in just 1 band. The former would serve as an anchor to calibrate systematics over a sufficiently large, representative area and the latter tier would provide the statistical constraining power. We further detail several options for the wide tier, e.g. utilizing the H-band, which would preserve the reference survey's dithering strategy or using the W-band, which has the advantage of a approximately four times higher survey speed compared to the H-band. While the latter would allow Roman to cover 13,000 deg$^2$ within one year of observations at the same depth as LSST, it would also require a stringent demonstration that wavelength-dependent PSF systematics are well controlled.

The tools presented here will be extended in several ways in future work. The framework should be extended to handle multiple survey layers with correlated systematics (e.g., to handle the increased uncertainty in dust extinction/reddening as one approaches the Galactic Plane; allowing noise bias uncertainties to be greater in regions of lower coverage; and of course including multiple surveys with partial overlap). A more realistic parameterization of the chromatic effects would be especially valuable in considering Option 3. Collecting a small test area in the W band early in the Roman mission is relatively fast (one can do 36 deg$^2$ in 1 day); this, as well as lessons learned from the Euclid experience (the Euclid VIS instrument also has a wide band), would give much higher confidence that we can accurately forecast this fastest option. This forecasting machinery will be invaluable as the community plans Roman and (eventually) its extended mission, as well as the next generation of ground-based CMB experiments.


The Euclid mission and its VIS band based shape measurements face a similar challenge ($\Delta \lambda/\lambda$ for VIS is similar to Roman W-band); cosmology results from Euclid as well as experience with PSF modeling in the Roman microlensing survey will be informative whether a W-band Roman survey is a feasible option as an extension to the Roman survey. Comparing Roman and Euclid we note that the two-tier concept of Roman is highly synergistic with the Euclid and LSST datasets. As currently planned, Euclid and LSST will only overlap for $\sim$ 8200 deg$^2$ and Euclid will not be able to match LSST depth in the later years of Rubin operations. Roman can cover the entire extragalactic footprint of LSST at the relevant depth, and in the area where it overlaps with Euclid, Roman's wavelength coverage begins where Euclid's ends. 

In summary, the concept of a two-tier Roman survey, composed of a medium  tier and a wide, single-band tier, combines exquisite systematics control and statistical constraining power. It represents a promising survey design for Roman as a standalone mission, and even more so when synergies and overlap with other contemporary datasets, optical and CMB, are included in the optimization process.

\section*{Data Availability}
The data underlying this article will be shared on reasonable request to the corresponding author.

\section*{Acknowledgments}
The authors thank Kaili Cao, Nihar Dalal, and Olivier Dor\'e for useful comments. This work was supported by NASA ROSES ATP 16-ATP16-0084, NASA 15-WFIRST15-0008, and Roman WFS 22-ROMAN22-0016 grants. Simulations in this paper use High Performance Computing (HPC) resources supported by the University of Arizona TRIF, UITS, and RDI and maintained by the UA Research Technologies department. 


\bibliographystyle{mnras}
\bibliography{references.bib}



\bsp	
\label{lastpage}
\end{document}